\documentclass[12pt]{article}
\usepackage{epsfig}
\usepackage{graphicx}
\usepackage{amsmath}
\usepackage{color}
\newcommand{\be}{\begin{equation}}
\newcommand{\ee}{\end{equation}}
\newcommand{\bea}{\begin{eqnarray}}
\newcommand{\eea}{\end{eqnarray}}
\newcommand{\norsl}{\normalsize\sl}
\newcommand{\norsc}{\normalsize\sc}

\newcommand{\nn}{\nonumber}

\def \ksl {k \kern-.45em{/}}
\def \lsl {l \kern-.45em{/}}
\def \ppsl {p \kern-.45em{/}}
\def \qsl {q \kern-.45em{/}}

\begin{document}

\begin{titlepage}

\title{ The real  photon 
structure functions  in massive parton model in NLO}
\author{
\norsc Norihisa Watanabe$^{a)}$\thanks{e-mail address: norihisa@post.kek.jp}, Yuichiro Kiyo$^{b)}$\thanks{e-mail address: ykiyo@juntendo.ac.jp}, Ken Sasaki$^{c)}$\thanks{e-mail address: sasaki@ynu.ac.jp}~  \\
\norsl a) National Laboratory for High Energy Physics(KEK) \\
\norsl  Tsukuba, Ibaraki 305-0801, JAPAN \\
\norsl b)~ Dept. of Physics,  Juntendo University \\
\norsl   Inzai,  Chiba 270-1695, JAPAN \\
\norsl c) Dept. of Physics,  Faculty of Engineering, Yokohama National
University \\
\norsl  Yokohama 240-8501, JAPAN \\
}

\date{}
\vspace{2cm}
\maketitle

\vspace{2cm}

\begin{abstract}
{\normalsize
We investigate the one-gluon-exchange  ($\alpha \alpha_s$) corrections to the   real photon 
structure functions $W_{TT} $, $W_{LT}$, $W_{TT}^{a} $ and $W_{TT}^\tau$ in the massive parton model. 
 We employ a technique based on the Cutkosky rules  and the reduction of 
Feynman integrals to master integrals.  
We show that a positivity constraint, which is derived from the Cauchy-Schwarz inequality, is satisfied among the unpolarized and polarized structure functions $W_{TT}$, $W_{TT}^a$ and $W_{TT}^\tau$ calculated up to 
the next-to-leading order in QCD.  
}
\end{abstract}

\begin{picture}(5,2)(-290,-550)
\put(2.3,-95){YNU-HEPTh-13-101}
\end{picture}

\thispagestyle{empty}
\end{titlepage}
\setcounter{page}{1}
\baselineskip 18pt

\section{Introduction}

Although a Higgs particle has been discovered at the CERN Large Hadron Collider (LHC)
~\cite{LHC}, we need to examine all of its properties with great accuracy to verify its full identity. For that purpose, the construction of a new $e^+e^-$ collider machine called the International Linear Collider (ILC)~\cite{ILC} is much anticipated. Even in the experiments at the ILC,  a detailed knowledge of the SM at high energies, especially based on QCD, is still important.  

It is well known that, in high energy $e^+e^-$ collision experiments, the cross section of the two-photon processes $e^+e^-\rightarrow e^+e^- + {\rm hadrons}$~ dominates over other processes such as the annihilation process $e^+e^-\rightarrow \gamma^* \rightarrow {\rm hadrons}$. The two-photon processes  at high energies provide a good testing ground for studying the predictions of QCD. 
In particular, the two-photon
process in which one of the virtual photon is very far off-shell~(large $Q^2\equiv -q^2$), while the other 
is close to the mass shell (small $P^2\equiv -p^2$),  can be viewed as a deep-inelastic electron-photon
scattering where the target is a photon rather than a nucleon~\cite{WalshBKT}.
In this deep-inelastic scattering of a photon target, we can study the photon structure 
functions, which are the analogs of the nucleon structure functions. When  polarized  beams are used in $e^+e^-$ collision experiments, we can 
get  information on  the spin structure of the photon. 

For a real photon ($P^2=0$) target, there  appear four structure functions: three unpolarized 
structure functions $F_2^\gamma(x,Q^2)$,  $F_L^\gamma(x,Q^2)$ and  $W_3^\gamma(x,Q^2)$, and one 
spin-dependent structure function $g_1^\gamma (x,Q^2)$, where $x=Q^2/(2p\cdot q)$.   The analysis of 
$F_2^\gamma$ and   $F_L^\gamma$ was first made in the parton model (PM)~\cite{WalshZerwas} and then investigated in perturbative QCD (pQCD). The leading order (LO) QCD contributions to $F_2^\gamma$ and   $F_L^\gamma$ were derived by Witten~\cite{Witten} and a few years later the next-to-leading order (NLO) corrections were calculated~\cite{BBetc}. The structure function $F_2^\gamma$ has been analyzed up to the next-to-next-to-leading order (NNLO)~\cite{MVV}. The QCD analysis of  
the polarized structure function $g_1^\gamma(x,Q^2)$ 
was performed in the LO~\cite{KS} and in the NLO~\cite{SV,GRS}.
For more information on the theoretical and experimental investigation of 
both unpolarized and polarized photon structure, see Ref.\cite{Krawczyk}. The photon structure functions 
of a virtual photon target ($P^2\ne 0$) have also been  analyzed  in pQCD. For more information 
on the study of the virtual photon structure functions $F_2^\gamma(x,Q^2,P^2)$, $F_L^\gamma(x,Q^2,P^2)$ 
and $g_1^\gamma(x,Q^2,P^2)$, see, for example, Ref.\cite{USU}.

So far in most of the QCD analyses of the photon structure functions,  all the active quarks have been treated as massless. At high energies the heavy charm and bottom  quarks  also contribute to the photon structure functions and their mass effects may not be neglected. In fact, the NLO QCD corrections due to heavy quarks have been 
calculated for the unpolarized  photon  structure functions  $F_2^\gamma$ and 
$F_L^\gamma$~\cite{SmithvanNeerven}. The heavy quark mass effects on the polarized  photon structure function $g_1^\gamma$ were analysed at NLO in QCD in Ref.\cite{GRS} by using the LO result of the massive PM.
Recently, we have investigated the heavy quark mass effects on $g_1^\gamma$ in the massive PM
at NLO in QCD and have found numerically  that the first moment of  $g_1^\gamma$ vanishes 
up to the NLO~\cite{G1NLO}.

In this paper we  investigate the four real photon structure functions $W_{TT}$, $W_{LT} $, $W_{TT}^{a}$ and $W_{TT}^\tau$ in the massive PM at NLO in QCD, and examine whether a positivity constraint~\cite{SSU} is satisfied among the unpolarized and polarized structure functions $W_{TT}$,  $W_{TT}^{a}$ and $W_{TT}^\tau$ at NLO.
The photon structure functions are 
 defined in the lowest order of the QED coupling constant 
$\alpha=e^2/4\pi$ and, in this paper,  they are of order $\alpha$. 

In the next section we discuss the photon structure functions. 
In Sec.\ref{Method}  we explain the method which we employed to calculate these 
structure functions in the massive PM at  NLO. In Sec.\ref{NumericalPlot} the NLO results for  $W_{TT}$, $W_{LT} $, $W_{TT}^{a}$ and $W_{TT}^\tau$ are given as a function of $x$ for both cases of charm and bottom quarks. We find that the positivity constraint among $W_{TT}$,  $W_{TT}^{a}$ and $W_{TT}^\tau$ is indeed satisfied for all the allowed $x$ region.
The final section is devoted to the conclusion.
In appendix the resummation formulae for the structure functions are given.

\section{Photon structure functions}

\begin{figure}
\begin{center}
\includegraphics[scale=0.7]{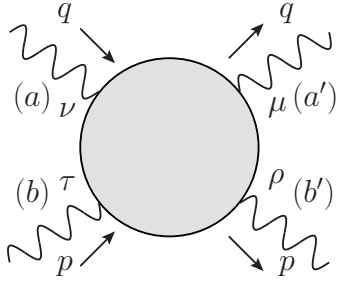}           
\caption{Photon-photon forward scattering with momenta $q(p)$ and helicities $a(b)$ and $a'(b')$  }
\label{Abspart}
\end{center}
\end{figure}

Let us consider the photon-photon forward  scattering amplitude, $\gamma(q,a)+\gamma(p,b)\rightarrow \gamma(q,a')+\gamma(p,b')$,  illustrated in figure~\ref{Abspart},
\bea
T^{\mu\nu\rho\tau}(p,q)=i\int d^4xd^4yd^4z e^{iq\cdot x}e^{ip\cdot(y-z) }\langle
 0\left|
T({\cal J}^\mu(x){\cal J}^\nu(0){\cal J}^\rho(y){\cal J}^\tau(z))
\right|0
\rangle,
\eea
where  $q$ and $p$ are four  momenta of the probe and target  photon, respectively, and ${\cal J}^\mu$ is the electromagnetic current. Its absorptive part is related to the structure tensor  
$W^{\mu\nu\rho\tau}$ as ~\cite{BCG} 
\bea
W^{\mu\nu\rho\tau}=\frac{1}{2\pi}{\rm Im}~T^{\mu\nu\rho\tau}.
\eea
 The s-channel helicity amplitudes are given by
\be
W(a'b'|ab)=\epsilon^*_{\mu}(a')\epsilon^*_{\rho}(b')W^{\mu\nu\rho\tau}\epsilon_{\nu}(a)\epsilon_{\tau}(b),
\ee
where 
 $\epsilon_{\mu}(a)$ represents
the photon polarization  vector, and $a,a'=0,\pm1$, and $b,b'=\pm1$. Note that the target photon is  real  and has no longitudinal mode. 
Due to the angular momentum conservation, parity conservation, and time reversal invariance, we have in total four independent s-channel helicity amplitudes\cite{BLS}, which we may take as
\bea
W(1,1|1,1), \quad W(0,1|0,1), \quad W(1,-1|1,-1), \quad W(1,1|-1,-1).
\eea
The first three amplitudes are helicity-nonflip and  the last one is helicity-flip.

For the real photon target, they appear four photon structure functions, 
$W_{TT}$, $W_{LT} $, $W_{TT}^{a}$ and $W_{TT}^\tau$, which are functions of $Q^2(=-q^2)$
and $x=Q^2/(2p\cdot q)$. They also depend on the active quark masses.
The subscripts $T$ and $L$ correspond to the transverse and longitudinal photon, respectively. The superscript $``a"$ 
of $W_{TT}^a $ refers to $\mu\nu$  antisymmentric part of $W^{\mu\nu\rho\tau}$, while the  superscript $``\tau"$
of  $W_{TT}^\tau $ refers  to transition with spin-flip for each of the photons. 
These structure functions are related to  the s-channel helicity amplitudes as follows;
\begin{subequations}
\bea
W_{TT}&=&\frac{1}{2}[W(1,1|1,1)+W(1,-1|1,-1)],\\
W_{LT}&=&W(0,1|0,1),\\
W_{TT}^a&=&\frac{1}{2}[W(1,1|1,1)-W(1,-1|1,-1)],\\
W_{TT}^\tau&=&W(1,1|-1,-1),
\eea
\end{subequations}
where 
$W_{TT}$,  $W_{LT} $ and  $W_{TT}^\tau$ are called as the unpolarized structure functions since they are measured,
for example, through the two-photon processes in unpolarized $e^+e^-$ collision experiments.  
When polarized $e^+$ and $e^-$ beams are used, we  can get information on the polarized structure function $W_{TT}^a$.
Other definitions of the photon structure functions are often used, which are $F_2^\gamma$, $F_L^\gamma$,  $g_1^\gamma$
and $W_3^\gamma$ and  are related to $W_i$'s as follows;
\bea
\begin{split}
F_2^\gamma=&2x\left[W_{TT}+W_{LT}\right],\quad
F_L^\gamma=2x W_{LT},\\
g_1^\gamma=&2W_{TT}^a,\quad W_3^\gamma=\frac{1}{2} W_{TT}^\tau.
\end{split}
\eea

There exist positivity constraints on the structure functions, which are derived from
 the Caushy-Schwarz inequality \cite{SSU} .
For the case of real photon, we obtain one positivity constraint   as follows;
\be
\left| W_{TT}^{\tau}\right| \leq \left|W_{TT}+W_{TT}^a\right|. \label{Positivity}
\ee
We will confirm positivity constraint at NLO.

\vspace{0.7cm}
\section{Calculation}\label{Method}
\begin{figure}
\begin{center}
\includegraphics[scale=0.65]{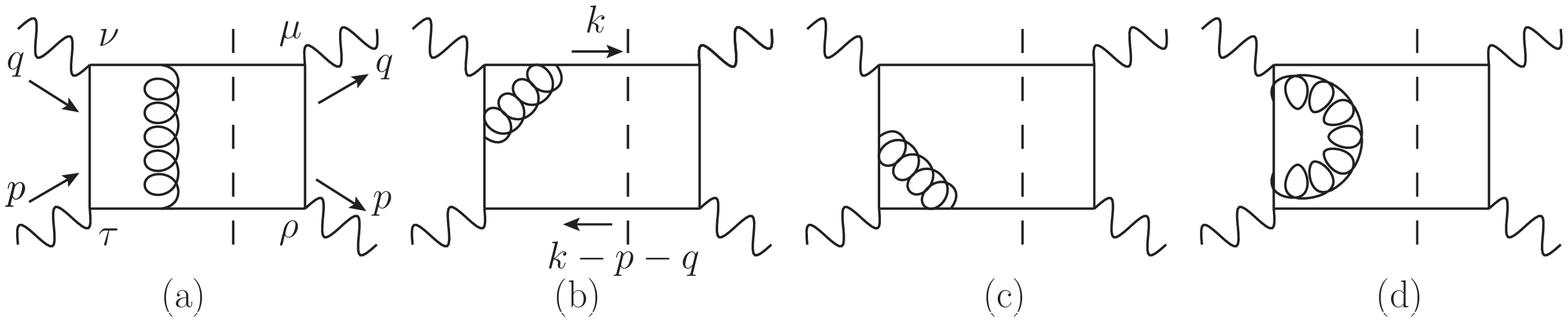}
\caption{Two-loop diagrams with virtual corrections. Graphs with virtual corrections to the right of the cut lines and graphs with $(q, \mu)$ and $(p,\rho)$ interchanged are added. Graphs with the external quark self-energies are not shown in the figure, but should be included in the 
calculation.}
\label{VirtualCorr}
\end{center}
\begin{center}
\includegraphics[scale=0.65]{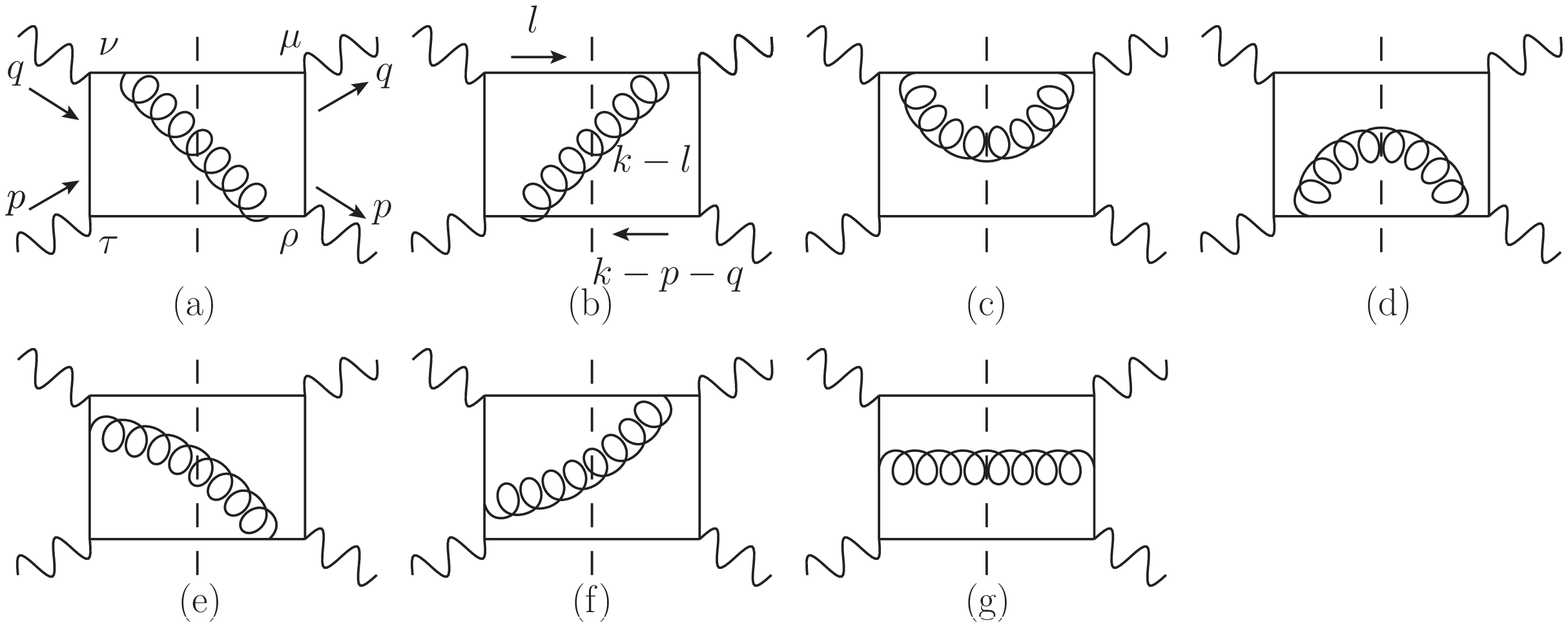}
\caption{Two-loop diagrams with a real gluon emission. Similar graphs corresponding to (e) and (f) are included. Also graphs with $(q, \mu)$ and $(p,\rho)$ interchanged are added. }
\label{RealEmis}
\end{center}
\end{figure}

We calculate the cross sections for 
the two photon annihilation to the heavy quark $q_H{\overline q}_H$ pairs 
\be
\gamma^*(q)+\gamma(p)\longrightarrow q_H+{\overline q}_H~, \label{BoxwithGluon}
\ee
with one-loop gluon corrections and to the gluon bremsstrahlung processes
\be
\gamma^*(q)+\gamma(p)\longrightarrow q_H+{\overline q}_H+g~.\label{Brems}
\ee
We employ the technique which is 
based on the Cutkosky rules ~\cite{AnaMel} and the reduction of Feynman integrals to master integrals.
First, following the Cutkosky rules \cite{Cutkosky}, the delta-functions which appear 
in the phase space integrals are replaced with differences of two propagators
\be
2\pi i\delta(r^2-m^2)\rightarrow \frac{1}{r^2-m^2+i0}-\frac{1}{r^2-m^2-i0}~,\label{CutkoskyRule}
\ee
where $m$ is the  quark mass.
Then the cross sections for the virtual corrections to the processes (\ref{BoxwithGluon}) and for the  bremsstrahlung processes (\ref{Brems}) are described by the two-loop diagrams shown in 
figure~\ref{VirtualCorr} and figure~\ref{RealEmis}, respectively, where a cut propagator should be understood as the r.h.s. of Eq.(\ref{CutkoskyRule}).

We regularize the amplitudes in  dimensional regularization $D=4-2\epsilon$. The absorptive part of the relevant
photon-photon scattering amplitude, $W^{\mu\nu\rho\tau}$,  can be written as \cite{BCG} 
\bea
W^{\mu\nu\rho\tau}&=&(T_{TT})^{\mu\nu\rho\tau}W_{TT}+(T_{LT})^{\mu\nu\rho\tau}W_{LT}
+(T_{TT}^a)^{\mu\nu\rho\tau}W_{TT}^a+(T_{TT}^\tau)^{\mu\nu\rho\tau}W_{TT}^\tau,\nn\\
\eea
where 
\begin{subequations}
\bea
(T_{TT})^{\mu\nu\rho\tau}&=&R^{\mu\nu}R^{\rho\tau},\\
(T_{LT})^{\mu\nu\rho\tau}&=&k_{1}^\mu k_1^\nu R^{\rho\tau},\\
(T^a_{TT})^{\mu\nu\rho\tau}&=&R^{\mu\rho}R^{\nu\tau}-R^{\mu\tau}R^{\nu\rho},\\
(T^\tau_{TT})^{\mu\nu\rho\tau}&=&\frac{1}{2}\left(
R^{\mu\rho}R^{\nu\tau}+R^{\mu\tau}R^{\nu\rho}-R^{\mu\nu}R^{\rho\tau}
\right),
\eea
\end{subequations}
with
\be
R^{\mu\rho}=-g^{\mu\rho}+\frac{q^{\mu} p^{\rho}+q^{\rho} p^{\mu}}{p\cdot q}-\frac{q^2p^{\mu} p^{\rho} }{(p\cdot q)^2},
\quad k_{1}^\mu=\sqrt{\frac{-q^2}{(p\cdot q)^2}} \left(p^{\mu}-\frac{p\cdot q}{q^2}q^\mu\right)~.
\ee

We introduce  the following $D$-dimensional projection operators
\begin{subequations}
\bea
&&\hspace{-1.cm}
(P_{TT})^{\mu\nu\rho\tau} = \frac{3D-8}{2D(D-2)(D-3)} (T_{TT})^{\mu\nu\rho\tau}+ 
\frac{D-4}{D(D-2)(D-3)}(T^\tau_{TT})^{\mu\nu\rho\tau}~,\\
 &&\hspace{-1.cm}
 (P_{LT})^{\mu\nu\rho\tau} = \frac{1}{D-2} (T_{LT})^{\mu\nu\rho\tau} ~,\\
&& \hspace{-1.cm}
(P_{TT}^a)^{\mu\nu\rho\tau} = \frac{1}{2(D-2)(D-3)} (T^a_{TT})^{\mu\nu\rho\tau} ~,\\
 &&\hspace{-1.cm}
 (P_{TT}^\tau)^{\mu\nu\rho\tau} = \frac{D-4}{D(D-2)(D-3)} (T_{TT})^{\mu\nu\rho\tau}+ 
\frac{2}{D(D-3)}(T^\tau_{TT})^{\mu\nu\rho\tau}~,
\eea
\end{subequations}
such that each  structure functions, $W_{TT}$,  $W_{LT} $, $W_{TT}^a$ and  $W_{TT}^\tau$, can be extracted
by means of the property
 $(P_i)^{\mu\nu\rho\tau} (T_j)_{\mu\nu\rho\tau} = \delta_{ij} $.

We apply the above  projection operators  to the two-loop diagrams given in figures~\ref{VirtualCorr}  and \ref{RealEmis}. 
 The contributions to each structure function are expressed in a linear combination of  two-loop scalar integrals of the form
\bea
&&\hspace{-1.5cm}A(\nu_i)
\equiv A(\nu_{k},\nu_{kq},\nu_{kp},\nu_{kpq},\nu_{l},\nu_{lq},\nu_{lp},\nu_{lpq},\nu_{kl})\nn\\
&&\hspace{-1cm}=\int\frac{d^D k}{(2\pi)^D}\frac{d^Dl}{(2\pi)^D}\frac{1}{D_{k}^{\nu_{k}}D_{k-q}^{\nu_{kq}}D_{k-p}^{\nu_{kp}}D_{k-p-q}^{\nu_{kpq}}D_{l}^{\nu_{l}}D_{l-q}^{\nu_{lq}}
D_{l-p}^{\nu_{lp}}D_{l-p-q}^{\nu_{lpq}}D_{k-l,0}^{\nu_{kl}}}~,\nn\\\label{As}
\eea
where 
\be
D_p=p^2-m^2,\qquad D_{p,0}=p^2.
\ee  
Note that  $ 1/D_{k-l,0}$ corresponds  to a gluon propagator.
 The coefficients of these scalar  integrals are written as functions of $x, Q^2, m^2$ and $D$. 
Actually  $A(\nu_i)$ has seven propagators at most and  at least two $\nu_i$'s are zero.
We choose the loop integration variables $k$ and $l$, such that  
momentum assignment of the cut propagators correspond to $1/D_{k}$ and $1/D_{k-p-q}$
 for the diagrams in figure~\ref{VirtualCorr} and 
$1/D_{l}$, $1/D_{k-p-q}$ and $1/D_{k-l,0}$  for the diagrams in figure~\ref{RealEmis}.
If  $\nu_i$'s of the cut propagators  are 0 or  negative integer, those integrals do not contribute to 
structure functions due to the Cutkosky rule. Thus we only pick up $A(\nu_i)$'s which are in the form 
$A(1,\nu_{kq},\nu_{kp},1,\nu_{l},\nu_{lq},\nu_{lp},\nu_{lpq},\nu_{kl})$ in figure~\ref{VirtualCorr}  and $A(\nu_{k},\nu_{kq},\nu_{kp},1,1,\nu_{lq},\nu_{lp},\nu_{lpq},1)$ in figure~\ref{RealEmis}. The other scalar integrals are  discarded.

There are still a large  number of  scalar integrals. Next, we apply the  reduction procedure  
\cite{Laporta} and rewrite the scalar integrals in terms of fewer number of master integrals.  This procedure  is based on the method of integration by parts 
\cite{Tkachov} and the use of the Lorentz invariance of  scalar integrals~\cite{GehrmannRemiddi}. 
We make use of {\tt FIRE}~\cite{FIRE}, a public reduction code powered by {\tt Mathematica}, and  express the relevant $A(\nu_i)$s as a linear combination of the master integrals, which are denoted as 
\be
M(\nu_i)\equiv M(\nu_{k},\nu_{kq},\nu_{kp},\nu_{kpq},\nu_{l},\nu_{lq},\nu_{lp},\nu_{lpq},\nu_{kl})~,
\ee
in the same way as the notation of $A(\nu_i)$s in Eq.(\ref{As}). Again the master integrals 
in the form of $M(1,\nu_{kq},\nu_{kp},1,\nu_{l},\nu_{lq},\nu_{lp},\nu_{lpq},\nu_{kl})$ are 
only relevant for the virtual-correction diagrams in figure.~\ref{VirtualCorr} and 
those in the form of $M(\nu_{k},\nu_{kq},\nu_{kp},1,1,\nu_{lq},\nu_{lp},\nu_{lpq},1)$ are relevant 
for the real-gluon-emission diagrams in figure.~\ref{RealEmis}. 

Finally, we perform the  phase space integrations  by taking  discontinuities of the  master integrals with cut propagators . 
For the two-cut and three-cut master integrals, we evaluate
\bea
&&\hspace{-1cm}{\rm Disc}^{(2)}~M(1,\nu_{kq},\nu_{kp},1,\nu_{l},\nu_{lq},\nu_{lp},\nu_{lpq},\nu_{kl})\nn\\
&&\hspace{-1cm}\equiv \int\frac{d^D k}{(2\pi)^{D}}(2\pi)\delta^{(+)}(k^2-m^2)(2\pi)\delta^{(+)}\left((p+q-k)^{2}-m^2\right)
\frac{1}{D_{k-q}^{\nu_{kq}}D_{k-p}^{\nu_{kp}}}\nn\\
&&\hspace{-0.5cm}\times\int\frac{d^Dl}{(2\pi)^D}\frac{1}{D_l^{\nu_{l}}D_{l-q}^{\nu_{lq}}
D_{l-p}^{\nu_{lp}}D_{l-p-q}^{\nu_{lpq}}D_{k-l,0}^{\nu_{kl}}}~,\nn\\
&&\label{PhaseSpaceInt2}\\
&&\hspace{-1cm}{\rm and}\nn\\
&&\hspace{-1cm}{\rm Disc}^{(3)}~M(\nu_{k},\nu_{kq},\nu_{kp},1,1,\nu_{lq},\nu_{lp},\nu_{lpq},1)\nn\\
&&\hspace{-1cm}\equiv \int \int\frac{d^D k}{(2\pi)^{D}}\int\frac{d^D l}{(2\pi)^{D}}~ 
(2\pi)\delta^+((k-l)^2)(2\pi)\delta^+(l^2-m^2)(2\pi)\delta^+((p+q-k)^2-m^2)\nn\\
&&\times\frac{1}{D_{k}^{\nu_{k}}D_{k-q}^{\nu_{kq}}D_{k-p}^{\nu_{kp}}
D_{l-q}^{\nu_{lq}}
D_{l-p}^{\nu_{lp}}D_{l-p-q}^{\nu_{lpq}}}~,\label{PhaseSpaceInt3}
\eea
respectively, and $M$'s are master integrals which remained after applying reduction algorithm. Note that  at least two 
$\nu_i$'s are zero in both (\ref{PhaseSpaceInt2}) and (\ref{PhaseSpaceInt3}).
The choice of a set of master integral is not unique and we are at liberty to replace a master integrals with one of other scalar
integrals. We choose a set of master integrals such that the coefficients of master integrals are finite in the limit 
$D\rightarrow 4$~\cite{ChetyrkinFST}.
With this choice of the set, 
the phase space integrations for  master integrals need only be evaluated  up to the finite terms in the series expansion in $\epsilon$. 

The ultraviolet (UV) singularities appear in the graphs (b), (c) and (d) of figure~\ref{VirtualCorr}, while 
the infrared (IR) singularities emerge from the graph (a) of figure~\ref{VirtualCorr} 
and from the real gluon emission graphs (a),(b), (c) and (d) of figure~\ref{RealEmis}.
  Both the UV and IR singularities are regularized 
by dimensional regularization. The UV singularities are removed by renormalization. 
We adopt the on-shell scheme both  for the wave function renormalization of the external 
quark and for the mass renormalization. For the latter, we  replace the bare mass in the Born cross section by the renormalized mass $m$, 
\be
m_{bare}\rightarrow m\left[1+\frac{\alpha_s(\mu)}{4\pi}C_F S^{\epsilon}\left(
\frac{\mu^2}{m^2}\right)^{\epsilon}\left\{-\frac{3}{\epsilon}-4\right\}\right]~,
\ee
where $\alpha_s(\mu)=g^2/(4 \pi)$ is the QCD running coupling constant, $C_F=4/3$ is the Casimir factor and $S^{\epsilon}=(4\pi)^\epsilon e^{-\epsilon \gamma_E}$ with 
Euler constant $\gamma_E$ and $\mu$ is the arbitrary reference scale of dimensional regularization. The renormalization of the QCD gauge coupling constant is not 
necessary at this order.
The IR singularities cancel when the both contributions from the virtual correction graphs and the real gluon
emission graphs are added. 
Actually the IR singularities reside in the two-cut master integrals in the form 
 $M(1,\nu_{kq},\nu_{kp},1,1,\nu_{lq},\nu_{lp},1,1)$ and the three-cut master integrals 
$M(\nu_{k},\nu_{kq},\nu_{kp},1,1,\nu_{lq},\nu_{lp},\nu_{lpq},1)$ with $\nu_{k}+\nu_{lpq}=2$.

\section{Numerical results}\label{NumericalPlot}

\begin{figure}[ht]
 \begin{tabular}{cc}
 \begin{minipage}{0.5\hsize}
  \begin{center}
   \includegraphics[scale=0.85]{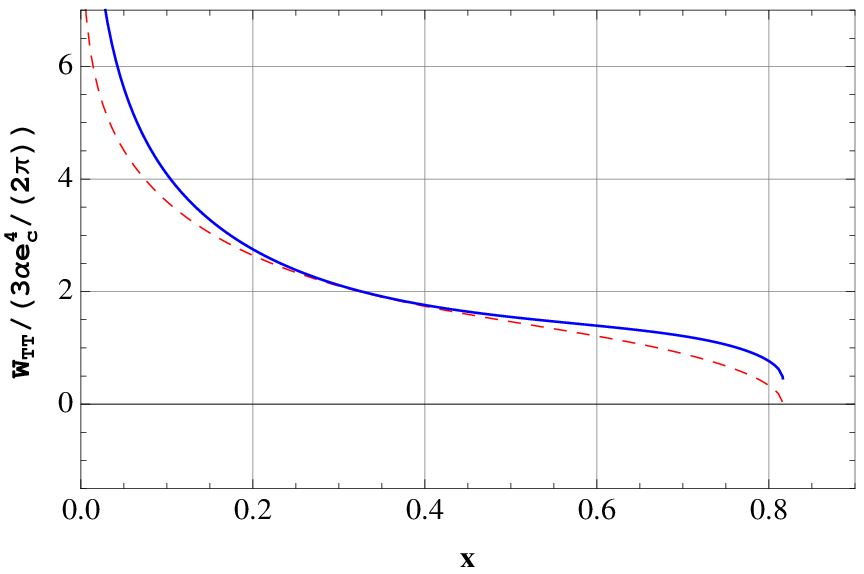}
   \hspace{1.6cm} (a)~$W_{TT}(x,Q^2)$
  \end{center}
 \end{minipage} &
 \begin{minipage}{0.5\hsize}
  \begin{center}
   \includegraphics[scale=0.85]{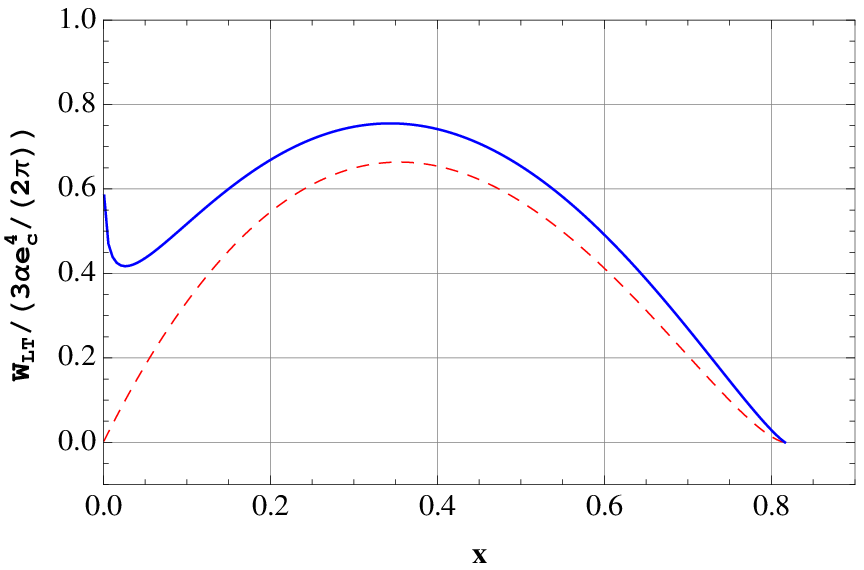}
   \hspace{1.6cm} (b)~$W_{LT}(x,Q^2)$
  \end{center}
 \end{minipage} \\
 \begin{minipage}{0.5\hsize}
  \begin{center}
   \includegraphics[scale=0.85]{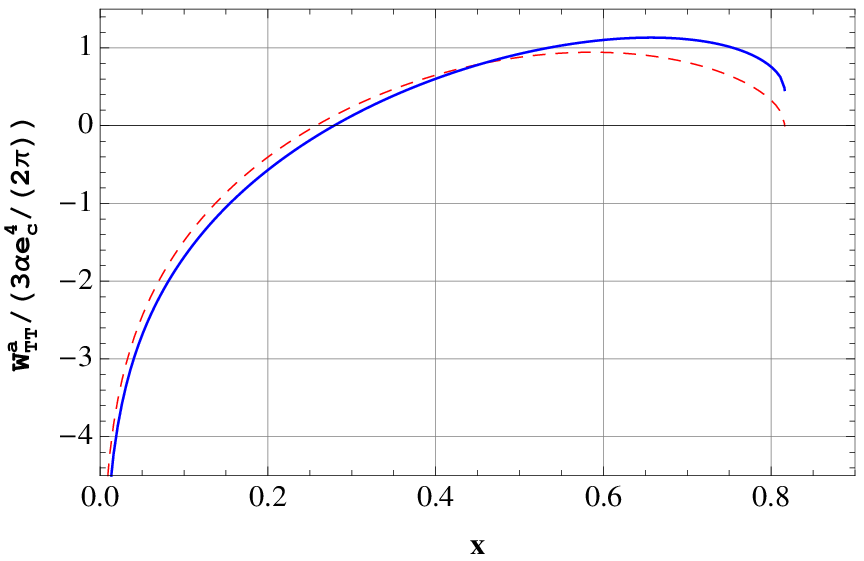}
   \hspace{1.6cm} (c)~$W_{TT}^a(x,Q^2)$
  \end{center}
 \end{minipage} &
 \begin{minipage}{0.5\hsize}
  \begin{center}
   \includegraphics[scale=0.85]{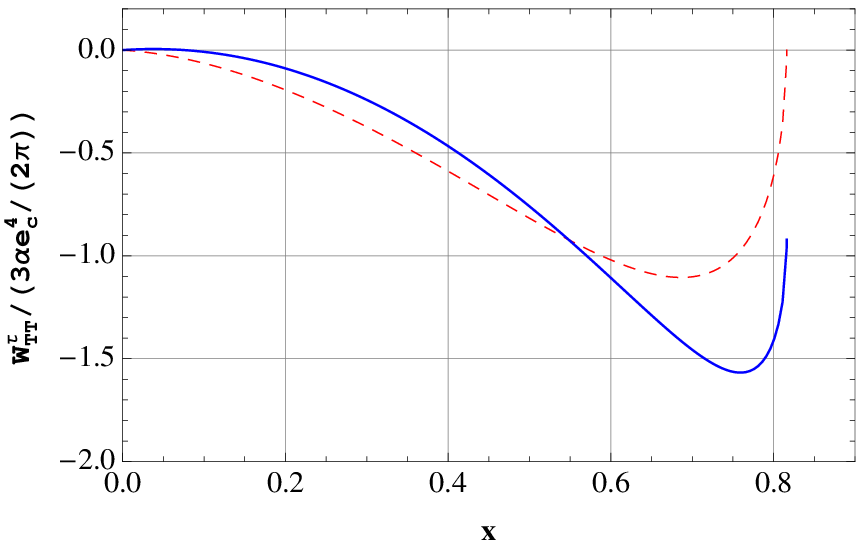}
   \hspace{1.6cm} (d)~$W_{TT}^\tau(x,Q^2)$
  \end{center}
 \end{minipage}
\end{tabular}
 \caption{The charm quark effects on the real photon structure functions (a) $W_{TT}$, (b) $W_{LT}$, (c) $W_{TT}^a$ and (d) $W_{TT}^\tau$, in the  PM    for $Q^2=30~ {\rm GeV}^2$,  
$m_c=1.3$ GeV and $e_c=\frac{2}{3}$ with $\alpha_s=0.21$. The vertical axes are in unit of $(3\alpha e_c^4/(2\pi))$. We plot the LO results (red dotted line) and
the  results up to the NLO (blue solid line).}
 \label{FigWic}
\end{figure}

\begin{figure}[ht]
 \begin{tabular}{cc}
 \begin{minipage}{0.5\hsize}
  \begin{center}
   \includegraphics[scale=0.85]{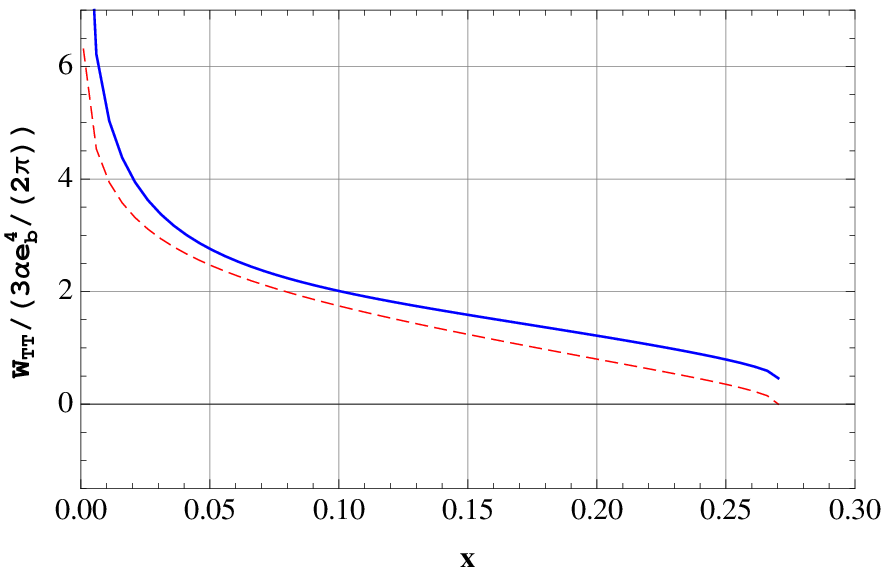}
   \hspace{1.6cm} (a)~$W_{TT}(x,Q^2)$
  \end{center}
 \end{minipage} &
 \begin{minipage}{0.5\hsize}
  \begin{center}
   \includegraphics[scale=0.85]{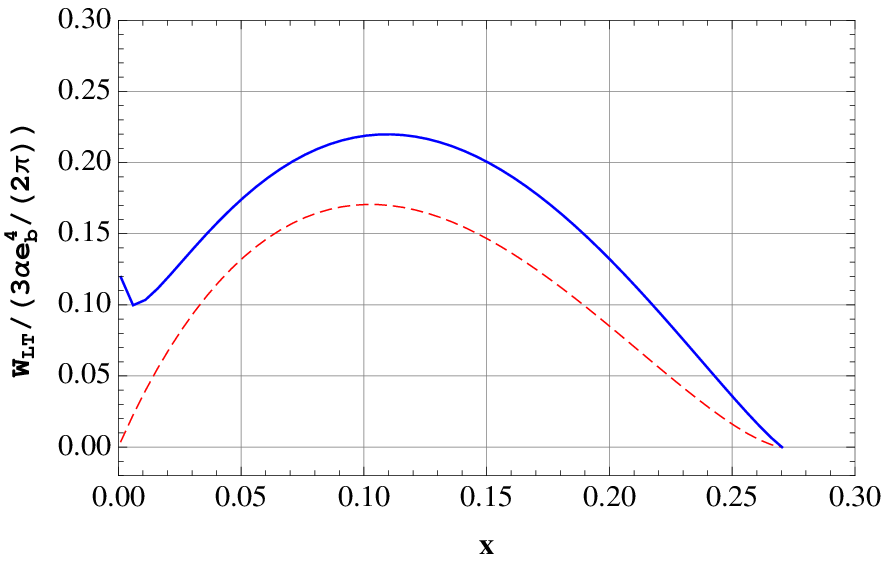}
   \hspace{1.6cm} (b)~$W_{LT}(x,Q^2)$
  \end{center}
 \end{minipage} \\
 \begin{minipage}{0.5\hsize}
  \begin{center}
   \includegraphics[scale=0.85]{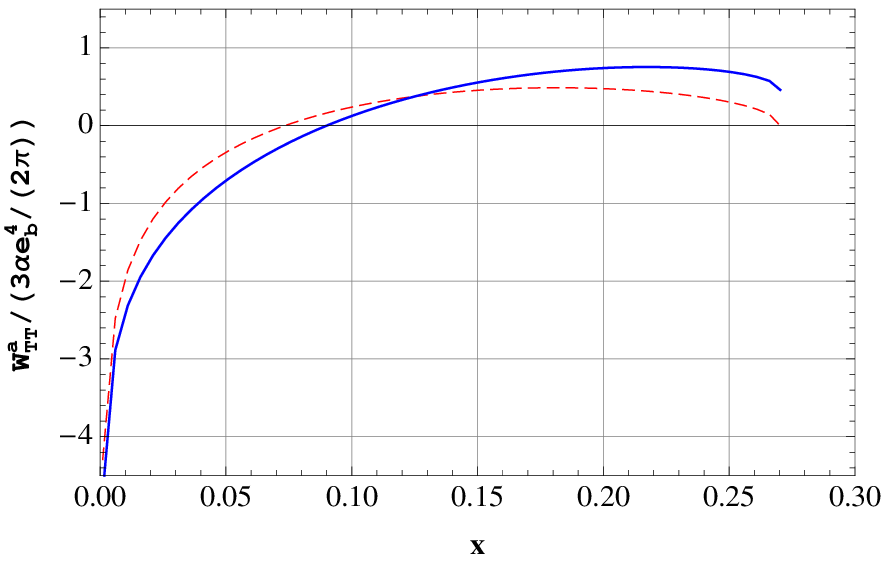}
   \hspace{1.6cm} (c)~$W_{TT}^a(x,Q^2)$
  \end{center}
 \end{minipage} &
 \begin{minipage}{0.5\hsize}
  \begin{center}
   \includegraphics[scale=0.85]{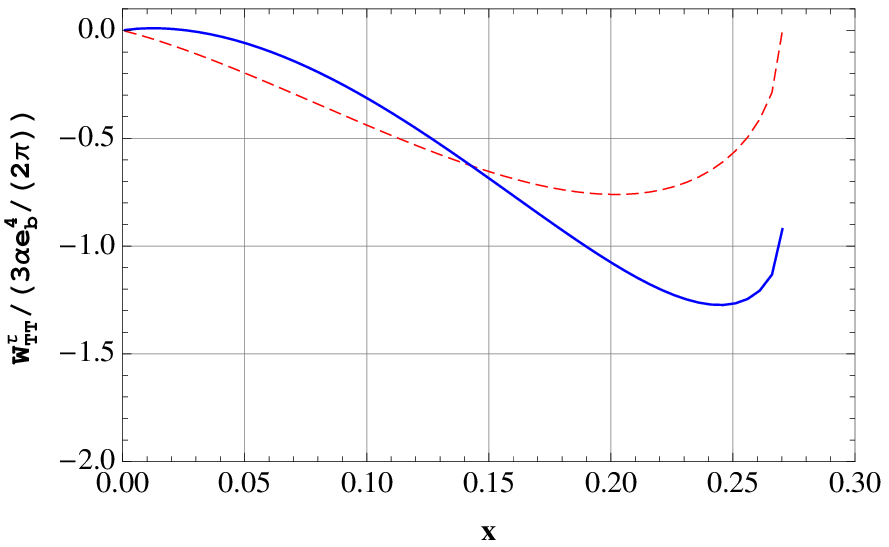}
   \hspace{1.6cm} (d)~$W_{TT}^\tau(x,Q^2)$
  \end{center}
 \end{minipage}
\end{tabular}
 \caption{The bottom quark effects on the real photon structure functions,  (a) $W_{TT}$, (b) $W_{LT}$, (c) $W_{TT}^a$ and (d) $W_{TT}^\tau$, in the PM  for $Q^2=30~ {\rm GeV}^2$,  
$m_b=4.5$ GeV and $e_b=-\frac{1}{3}$ with $\alpha_s=0.21$. The vertical axes are  in unit of $(3\alpha e_b^4/(2\pi))$.
We plot the LO results (red dotted line) and
the results up to the NLO (blue solid line).}
 \label{FigWib}
\end{figure}

  We plot  in figures.~\ref{FigWic} and \ref{FigWib} the  real photon   structure functions $W_{TT}(x,Q^2)$, $W_{LT}(x,Q^2)$, $W_{TT}^a(x,Q^2)$  and $W_{TT}^\tau(x,Q^2)$   predicted by the massive PM up to the NLO for the case of $Q^2=30~ {\rm GeV}^2$. We choose  charm and bottom quark  as a 
heavy quark for the  figures~\ref{FigWic} and \ref{FigWib}, respectively.
For the running coupling constant, we take $\alpha_s=0.21$ choosing $\mu^2=Q^2$.  
We take   $m_c=1.3~ {\rm GeV}$, $m_b=4.5~ {\rm GeV}$ , $e_c=\frac{2}{3}$ and $e_b=-\frac{1}{3}$. 
 We show two curves for each structure function:  the LO result  and the result up to the NLO.   The allowed $x$ region is $0\le x \le x_{Q}$ ~with  
\be
x_{Q}=\frac{1}{1+\frac{4m^2}{Q^2}}~.
\ee
The LO results are already known \cite{BCG}
such as
\begin{subequations}
\bea
W_{TT}^{\rm LO}(x,Q^2)&=&\delta_q ~
\Bigl\{-\frac{1}{2}\Bigl(\ln\frac{1+\beta}{1-\beta}\Bigr)\left((\beta^4-5)x^2-2(\beta^4-3)x+\beta^4-3\right)\nn\\
&&\qquad\quad
+\beta\left((\beta^2-5)x^2-2(\beta^2-3)x+\beta^2-2\right)
\Bigr\},\\
W^{\rm LO}_{LT}(x,Q^2)&=&
\delta_q ~2(1-x)x\Bigl\{\Bigl(\ln\frac{1+\beta}{1-\beta}\Bigr)(\beta^2-1)+2\beta\Bigr\},\\
W_{TT}^{a,\rm LO}(x,Q^2)
&=&
\delta_q  \Bigl\{\Bigl(\ln\frac{1+\beta}{1-\beta}\Bigr)(2x-1)+\beta(-4x+3)\Bigr\},
\\
W_{TT}^{\tau,\rm LO}(x,Q^2)&=&
\delta_q ~
\Bigl\{-\frac{1}{2}\Bigl(\ln\frac{1+\beta}{1-\beta}\Bigr)(\beta^2-1)(x-1)\left((\beta^2+3)x-\beta^2+1\right)\nn\\
&&\qquad\quad
+\beta\left((\beta^2-3)x^2-2(\beta^2-1)x+\beta^2-1\right)
\Bigr\},
\eea
\end{subequations}

where 
\be
\beta=\sqrt{1-\frac{4m^2 x}{Q^2(1-x)}} =\sqrt{\frac{1-\frac{x}{x_Q}}{1-x}},
\ee
and 
\be
\delta_q = \frac{3\alpha}{2\pi}e_q^4.
\ee
For $x\rightarrow x_{Q}$, $\beta$ goes to zero and thus structure functions vanishes at $x_{Q}$ in LO. 

We  observe  that the radiative corrections in NLO are noticeable.
In the graphs (a), (c) and (d) of 
 figures.~\ref{FigWic} and \ref{FigWib} we find that
the radiative corrections to $W_{TT}$, $W_{TT}^a$ and $W_{TT}^\tau$ are large near the threshold 
(near $x_{Q}$). Indeed those NLO curves do not vanish at $x_{Q}$. This is due to the  Coulomb 
singularity, which appears when the Coulomb gluon is exchanged between the quark and anti-quark pair near threshold. The  diagram figure~\ref{VirtualCorr}(a) is responsible for this threshold behaviour.
The virtual correction to the left of the cut line  in figure~\ref{VirtualCorr}(a) gives rise to a factor  $1/\beta$ while 
a factor $\beta$ comes out from the phase space integration. They are combined and yield a finite  but non-zero result at $x_{Q}$. 
On the other hand,
 the  radiative corrections to $W_{LT}$  shown in  the figures ~\ref{FigWic}(b) and \ref{FigWib}(b) evade the Coulomb singularity  and vanish at threshold.   This is because  the  structure  of Coulomb enhancement is given by  
 $W_{i}^{\rm NLO}\sim \left(W_{i}^{\rm LO}\right)\times \alpha_s/\beta$ and $W_{LT}^{\rm LO}$ behaves as $\beta^3$ for $\beta\rightarrow 0$. Thus  $W_{LT}^{\rm NLO}$ vanishes as $\beta^2$ near the threshold.
 
   The jump size of Coulomb enhancement for each structure function  is calculable in another
way.   
It is well known that the contributions of Coulomb gluons can be summed up to all order. The result of all order resummation 
 is given by Sommerfeld factor.  Using Taylor expansion of Sommerfeld  factor 
  in strong coupling constant $\alpha_s$, leading Coulomb singularity can be reproduced to all order in $\alpha_s$.
   Combining the LO photon structure function $W^{\rm LO}$ with the Sommerfeld factor, we 
can predict that the jump of the structure function are given by a derivative of LO structure
function at the threshold multiplied by $C_F \alpha_s \pi/2$. That is given by 
\begin{eqnarray}
W_i^{\rm NLO}(x,Q^2) 
&\stackrel{\beta\to 0}{=}&
\left(\beta \left[\frac{d W_i^{\rm LO}}{d\beta}\right]_{\beta=0} 
\right)\times  \left( \frac{C_F \alpha_s \pi}{2\beta}\right)
+{\cal O}(\beta),\label{NLOjump}
\end{eqnarray}
where part in the first parenthesis  corresponds to phase space integration of heavy 
quark pair and squared LO amplitude and the second is the Coulomb singularity. 
This formula is assured by the factorization of hard correction and the Coulomb 
singularities near heavy quark threshold. 
Our  NLO results   are consistent with (\ref{NLOjump}) near the threshold.  
The formula also predict that the NNLO calculation in the massive PM 
suffers from a divergence,  $\beta \times (C_F \alpha_s/(2\beta))^2$, near threshold 
due to double Coulomb gluon exchange. Therefore fixed order calculation 
near threshold becomes ill-defined and we need to resort to the method of 
resummation of the Coulomb 
singularities.  Resummation formulae for the structure functions are given in  Appendix~{\ref{SecResum}}.

\begin{figure}[htbp]
  \begin{center}
    \begin{tabular}{c}

      \begin{minipage}{0.5\hsize}
        \begin{center}
          \includegraphics[scale=0.9]{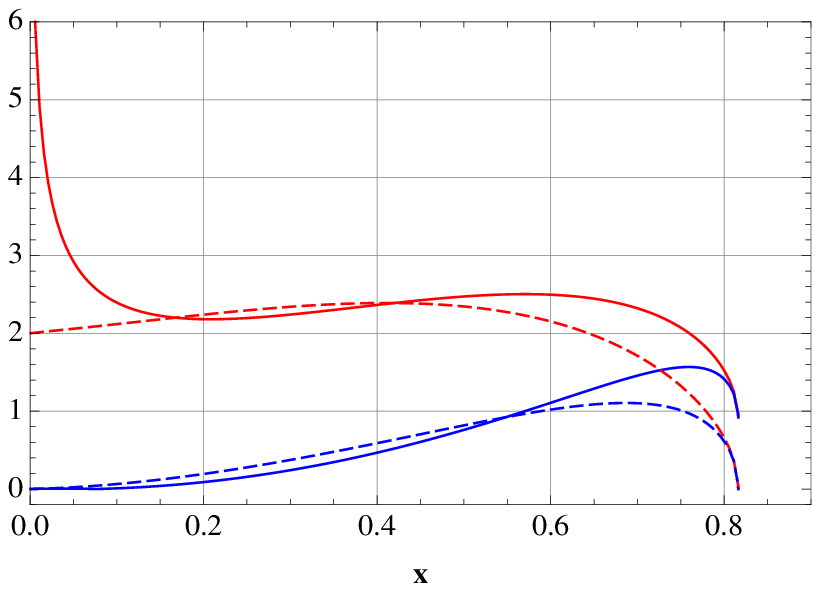}
          \hspace{1.6cm} (a)~charm quark
        \end{center}
      \end{minipage}

      \begin{minipage}{0.5\hsize}
        \begin{center}
          \includegraphics[scale=0.9]{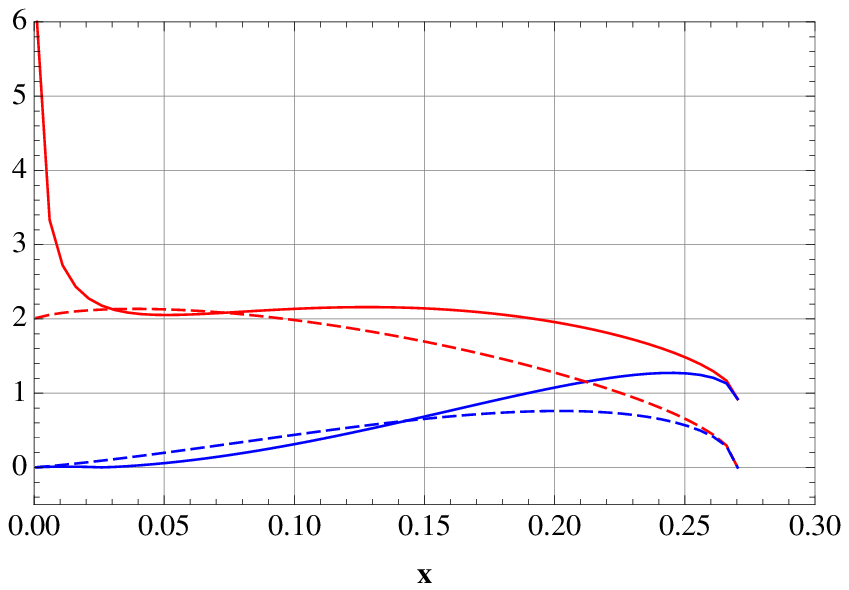}
          \hspace{1.6cm} (b)~ bottom quark
        \end{center}
      \end{minipage}

    \end{tabular}
    \caption{ The positivity constraint and the PM prediction up to the NLO for $Q^2=30~ {\rm GeV}^2$ and $\alpha_s=0.21$. (a) Case of charm quark with $m_c=1.3$ GeV and $e_c=\frac{2}{3}$; (b) Case of bottom quark with $m_b=4.5$ GeV and $e_b=-\frac{1}{3}$. The vertical axes are  in unit of $(3\alpha e_b^4/(2\pi))$. The result up to the NLO (LO)  for ($W_{TT}+W_{TT}^a$) is depicted in red solid (red dashed) line. The result up to the NLO (LO) for $|W_{TT}^\tau|$ is depicted in blue solid (blue dashed) line.}
    \label{ResPositiv}
  \end{center}
\end{figure}

 For $x\rightarrow 0$, the NLO contributions to $W_{TT}$ and $W_{TT}^a$ both diverge. The sum $(W_{TT}+W_{TT}^a)$ remains finite in  LO but diverges in  the NLO as $x\rightarrow 0$ [see figure.~\ref{ResPositiv}(a) and (b)]. This is due to the collinear divergence.  The limiting procedure 
$x=\frac{Q^2}{s+Q^2}\rightarrow 0$ with fixed $Q^2$ is equivalent of taking $s\rightarrow \infty$. Thus the situation at $x=0$ is the same as if we are dealing with massless quarks. When a gluon is 
emitted from a massless quark, a collinear divergence appears. We also see in figure.~\ref{FigWic}(b) the rise of the NLO 
contributions to $W_{LT}$ near $x=0$, which is again due to the collinear divergence. 
On the contrary, the LO and NLO contributions to $W_{TT}^\tau$ vanishes at $x=0$.  A collinear divergence does not occur for the helicity-flip amplitude $W_{TT}^\tau$.


 We plot the PM predictions of ($W_{TT}+W_{TT}^a$) and $|W_{TT}^\tau|$ for the case of charm quark 
in  figure~\ref{ResPositiv}(a) and for the bottom  case in figure~\ref{ResPositiv}(b).
In both cases we observe that the positivity constraint 
(\ref{Positivity}) for a real photon target is  satisfied up to the NLO for all the allowed $x$ region with a wide 
margin except at the threshold $x_{Q}$. At the threshold, we can find the following relation;
\bea
   |W_{TT}^\tau(x_{Q},Q^2) | =W_{TT}(x_{Q},Q^2)+W_{TT}^a(x_{Q},Q^2),\label{RelationATxmax}
\eea
At the threshold, we find the following relation  (\ref{RelationATxmax})
from our numerical analysis. This is also checked analytically 
using resummation formula given in the Appendix \ref{SecResum}.

 \section{Summary}
In this  paper, we have investigated heavy quark mass effects for the  real photon structure functions in the massive PM in the NLO in QCD. 
There are four structure functions  $W_{TT}$, $W_{LT}$, $W_{TT}^a$  and $W_{TT}^\tau$ for the real photon target. 
We have found that the radiative corrections to $W_{TT}$, $W_{TT}^a$ and $W_{TT}^\tau$   are large near the threshold. This is due to the  Coulomb 
singularity, which appears when the Coulomb gluon is exchanged between the quark and anti-quark pair near threshold.  On the other hand,  the  radiative corrections to $W_{LT}$  evade the Coulomb singularity  and vanish at threshold.   
We also have found that although the sum $(W_{TT}+W_{TT}^a)$ remains finite in  LO but diverges in the  NLO as $x\rightarrow 0$.  This is due to the collinear divergence.  The limiting procedure 
$x=\frac{Q^2}{s+Q^2}\rightarrow 0$ with fixed $Q^2$ is equivalent of taking the high energy limit $s\rightarrow \infty$. In other words, the situation at $x=0$ corresponds to the massless quark limit. A collinear divergence appears when a gluon is emitted from a massless quark. We also see the collinear divergence in the NLO 
contributions to $W_{LT}$ near $x=0$. 
Finally we have shown from the numerical plots of the PM predictions of ($W_{TT}+W_{TT}^a$) and $|W_{TT}^\tau|$  and the positivity constraint 
(\ref{Positivity}) for a real photon target is satisfied  for all the allowed $x$ region.




 \appendix{
\section{Threshold resummation for structure functions}\label{SecResum}
The Green function sums up 
leading Coulomb singularity for photon-photon forward scattering amplitude. The contribution 
to the structure function is given by  imaginary part of Green function
\begin{eqnarray}
{\rm Im}\,
G_C(\beta)
=
\frac{m^2 \beta}{4\pi}\bigg[
\frac{\frac{C_F \alpha_s \pi}{\beta} }{1-e^{-\frac{C_F\alpha_s \pi}{\beta}}}   \theta(\beta)
+ 
\frac{4\pi}{\beta}
\sum_{n=1}^\infty 
a_n^3 \, \delta \left(\beta^2+a_n^2 \right)
\bigg],
\end{eqnarray}
with $a_n =C_F\alpha_s/(2n)$.  The terms with $\delta$-function are due
 to Coulomb bound-states, which can have non-zero contribution to the structure functions  
 for $s < 4m^2$ because $\beta=i \sqrt{4m^2/s-1}$ becomes pure imaginary.
 
We combine the LO photon  structure function $W^{\rm LO}$ and the Coulomb 
Green function $G(\beta)$~\cite{HSY,Kiyo} in the following form 
 \begin{eqnarray}
 \widehat{W}_i^{\rm LO} (x,Q^2) 
 &=&
 W_i^{\rm LO} (x,Q^2) S(x) +  C_i\, T(x), 
 \end{eqnarray}
 where $S$ encodes the Coulomb singularity and $T$ 
 is the contribution due to boundstate poles. They are defined by   
 \begin{eqnarray}
 S(x)
 &=&
\frac{ C_F \alpha_s \pi \sqrt{\frac{x_Q(1-x)}{x_Q-x} } 
     }{1-\exp\left\{-C_F\alpha_s \pi \sqrt{\frac{x_Q(1-x)}{x_Q-x}}\right\} } 
 \theta(x_{Q}-x),
\\
T(x) &=& 
\sum_{n=1}^{\infty}  
\frac{4\pi  a_n^3 x_Q(1-x_Q)}{(1+a_n^2 x_Q)^2}
\delta \left(\frac{x_Q(1 +a_n^2)}{1+a_n^2 x_Q}  -x\right).
 \end{eqnarray} 
The matching factors are calculable for each structure function as 
 \bea
 C_i =\left\{C_{TT}~,C_{TT}^a,~C_{TT}^\tau\right\}=\left\{1, 1, -2 \right\}\delta_q.
 \eea
  The resummation formula for structure function can be applied for the cases 
  $W_{TT}$, $W_{TT}^a$ and $W_{TT}^\tau$.
Near threshold $W_{LT}$ is order of $\beta^3$ at LO, which is suppressed by
$\beta^2$ compared to $S$-wave case. Therefore its Coulomb singularity at NLO
is suppressed by $\beta^2$ and the resummation effect becomes moderate.





\end{document}